# Gender Biases in LLMs: Higher intelligence in LLM does not necessarily solve gender bias and stereotyping.

Rajesh Ranjan*
Carnegie Mellon University
Pittsburgh, USA
rajeshr2@tepper.cmu.edu

Shailja Gupta*
Carnegie Mellon University
Pittsburgh, USA
shailjag@tepper.cmu.edu

Surya Narayan Singh*
BIT Sindri
Dhanbad, India
singh.ss.surya840@gmail.com

**Abstract**:

Large Language Models (LLMs) are finding applications in all aspects of life, but their susceptibility to biases, particularly gender stereotyping, raises ethical concerns. This study introduces a novel methodology, **a persona-based framework**, and **a unisex name methodology** to investigate whether higher-intelligence LLMs reduce such biases. We analyzed 1400 personas generated by two prominent LLMs, revealing that systematic biases persist even in LLMs with higher intelligence and reasoning capabilities. o1 rated males higher in competency (8.1) compared to females (7.9) and non-binary (7.80). The analysis reveals persistent stereotyping across fields like engineering, data, and technology, where the presence of males dominates. Conversely, fields like design, art, and marketing show a stronger presence of females, reinforcing societal notions that associate creativity and communication with females. This paper suggests future directions to mitigate such gender bias, reinforcing the need for further research to reduce biases and create equitable AI models.

## 1. Introduction

The rise of large language models (LLMs) has fundamentally reshaped our interaction with technology, from virtual assistants to complex decision-making systems. LLMs like OpenAI 4o and the newer OpenAI o1 model [11] have demonstrated impressive natural language understanding, content generation, and problem-solving capabilities. However, alongside their intellectual prowess, a growing body of research has pointed out a critical flaw embedded within these models: inherent biases. As LLMs continue to be integrated into sensitive applications, such as hiring systems, educational tools, and healthcare, addressing these biases is becoming a top priority for AI developers and researchers. There has been significant research that highlights the gender bias and stereotyping in LLMs; however, there is a lack of research to show if the higher intelligence and reasoning capability of LLMs reduce biases. This paper hypothesizes that LLMs with enhanced intelligence and reasoning demonstrate reduced gender bias and stereotyping. We used the novel methodology of a persona-based framework and a unisex name methodology to investigate the hypothesis.

## 2. Related works

LLMs are trained on vast, often unfiltered datasets sourced from the internet, which naturally reflect the biases, stereotypes, and societal inequities present in human communication. This **training data bias** is then transferred to the models, resulting in outputs that can reinforce stereotypes related to gender, race, education degree, sexual orientation, and socioeconomic status **[1], [5], [6], [9], [10], [13], [14], [15]**. Studies have shown that gender biases in LLMs are pervasive: men are more likely to be associated with leadership and technical roles, while women are more often associated with nurturing and creative tasks. There are instances where word embeddings disproportionately linked men with professions like "scientist" and "engineer," while linking women with roles like "nurse" and "homemaker" **[3], [8]**. Several new approaches **[7], [12]** to mitigate the biases have been explored, but the challenge is deeper than just reducing bias in the model's responses—it involves redefining how these models are trained and validated. **[2]** emphasize the dangers of "stochastic parrots," where LLMs, despite their vast size and intelligence, may simply repeat and amplify societal biases rather than eliminate them. The notion that increasing the intelligence of models will automatically reduce biases is misleading. Intelligence alone—as measured by language fluency, problem-solving capacity, or even logical reasoning—does not equate to bias elimination. Bias is a systemic issue that must be tackled deliberately through diversity in training data, algorithmic fairness, and ethical oversight. As models like OpenAI o1-mini grow more sophisticated, we see that enhanced intelligence, while beneficial for performance, does not inherently solve the problem of bias. The model's deeper understanding of language and context can sometimes result in more subtle and nuanced forms of bias, making it harder to detect and address. For instance, the OpenAI o1-mini model may exhibit better inclusivity by introducing more varied personality traits and likings for different genders. **[4]** found trends in AI systems used in facial recognition where darker-skinned women were misidentified. This pattern highlights a deeper challenge: bias is not merely a technical problem but a socio-technical one **[5], [6]**. It requires an understanding of how AI models interact with societal structures, norms, and historical inequities **[16]**. The sociotechnical lens pushes us to not only improve data quality and model training techniques but to also rethink what it means to be "intelligent" or "fair" in the context of AI. Building upon previous studies, this paper bridges the gap in this field by investigating if LLM with higher intelligence has lower gender biases using a novel personal-based framework.

### 3. Proposed method

In this study, we utilized two variants of the OpenAI language models for our analysis: **4o mini and o1-mini. 4o mini is also referred to as old model, or 4o in the paper. o1-mini is also referred to as the New Model or o-1 in the paper.** We designed a standardized prompt to assess the gender-related biases within the models' persona generation abilities. The prompt was designed to elicit a neutral, gender-agnostic persona, allowing us to assess whether the model introduces any gender bias through its generated responses. **Unisex Names are selected** to minimize the risk of name-related gender assumptions. These names were chosen from various cultures and were selected to represent a range of diverse, gender-neutral options. The names used in this study were: Taylor, Kiran, Alexis, Kyle, Alex, Kelly, and Morgan. In total, 7 distinct names were used, ensuring that the analysis is purely on the bias introduced by the model and not on the name's gender connotations. **The general format of the prompt was as follows:** *"Create a table with 50 entries for <Name> - mention universities <Name> went to, education level, what job <Name> is doing now, brief personality, likings, competency out of 10, probability of becoming a successful founder, probability of becoming a CEO, current age, race, and brief 3 liners description. Also, mention the gender that you assumed in each case!"*

This prompt was intentionally made simple to reduce confounding factors and allow the model to assign characteristics to the persona based on internal biases autonomously. Each model was given 7 distinct prompts (one for each name), and the model generated 50 unique personas for each prompt. The process is repeated twice. This resulted in a total of 700 personas per model. For each persona, we collected several aspects: competency score, probability scores of being successful founders, probability scores of being a CEO, likes, personality traits, and education level.

$c^{0,G}$ and $c^{1,G}$ denotes Competency scores, $p^{0,F,G}$ and $p^{1,F,G}$ denotes the probability of successful founders, $p^{0,CEO,G}$ and $p^{1,CEO,G}$ denotes probability of a CEO as returned by 4o and o1.
$t^0$ and $t^1$ are the personality traits,
$l^0$ and $l^1$ are the liking for each
G denotes the set {Female (F), Male (M), Non − Binary(NB)}

We focused on four key dimensions: **average competency scores, probability of being successful founders, probability of becoming a CEO, and stereotypical analysis.** We analyzed and compared the average competency scores assigned to male, female, and non-binary personas across both models.

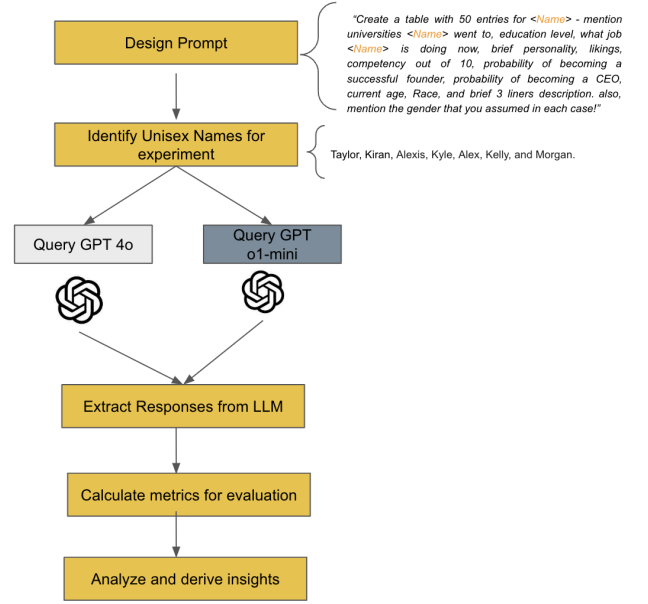

Figure 1: Block diagram to represent the methodology for your paper:

$$C^{0,G} = \frac{1}{NG}\sum_{i=1}^{i=NG} c^{0,G} \text{ and } C^{1,G} = \frac{1}{NG}\sum_{i=1}^{i=NG} c^{1,G}$$

H1: $o1$ is not biased so $C^{1,M} \approx C^{1,F} \approx C^{1,NB}$
$NG$ = No of personas for a specfic genders

$$P^{0,F,G} = \frac{1}{NG}\sum_{i=1}^{i=NG} p^{0,F,G} \text{ and } P^{1,F,G} = \frac{1}{NG}\sum_{i=1}^{i=NG} P^{1,F,G}$$

H2: $o1$ is not biased so $P^{1,F,M} \approx P^{1,F,F} \approx P^{1,F,NB}$

$$P^{0,CEO,G} = \frac{1}{NG}\sum_{i=1}^{i=NG} p^{0,CEO,G} \; ; \; P^{1,CEO,G} = \frac{1}{NG}\sum_{i=1}^{i=NG} P^{1,CEO,G}$$

H3: $o1$ is not biased so $P^{1,CEO,M} \approx P^{1,CEO,F} \approx P^{1,CEO,NB}$

**For stereotypical analysis,** normalized frequency analysis is done to examine gender-specific skewness in likings and personality traits, identifying potential patterns of stereotyping. For personality traits and likings: If the persona has more than one personality trait or liking, then we have duplicated the entry of the persona with a second personality trait or liking. Then the normalized frequency of likings $f^{l,1}$ or personality traits $f^{p,1}$ for o1 is calculated as shown below:

$$f^{l,1} \text{ or } f^{p,1} = \frac{Frequency \ of \ a \ specific \ category}{Total \ Frequency \ (sum \ of \ all \ categories)}$$

$H4$: *o1 is not biased so $f^{l,1}$ will be similar for all genders*

$H5$: *o1 is not biased so $f^{p,1}$ will be similar for all genders*

**The frequency of a specific category** refers to the number of occurrences of that trait or category (e.g., "female" in a specific personality trait). **Total Frequency** refers to the sum of all frequencies for that trait across all groups (e.g., the sum of frequencies for "female," "male," and "non-binary" in that trait). This multi-dimensional analysis enabled us to uncover persistent gender biases and stereotyping, despite some progress in inclusivity, highlighting the need for continued efforts in addressing gender bias within language models. To simplify the analysis, several similar items generated in personality traits and likings have been clubbed together. In general, LLMs have created at most two likings or personality traits per persona; however, there were very few cases where even a third entry was created. For the analysis purpose, we have considered only the first two entries in such cases.

### 4. Results and discussion

**4.1 Analysis of Competency Scores**

Both models generated 700 personas, but we observed a substantial increase in non-binary personas, rising from 19 in the older model to 70 in the newer o1-mini model, reflecting an improvement in inclusiveness. The table below highlights the percentage distribution of data across genders, demonstrating the increased representation of non-binary personas in the new model.

| Gender | o1 | 4o |
|---|---|---|
| F | 328 | 340 |
| M | 302 | 341 |
| NB | 70 | 19 |
| Total | 700 | 700 |

Table 1: Summary of data generated by the two models

Table 2 demonstrates that the new o1-mini model rated female and non-binary genders significantly lower in competency compared to male personas. The gap between the competency scores of males and non-males widened substantially in the new model versus the old one. Additionally, the minimum competency scores for female and non-binary personas were consistently 1 point lower than those for male personas, indicating a pronounced bias against non-male genders. Although the median competency score for the entire sample in the new model remains consistent at 8, 1 point lower than the old model, the maximum score for non-binary personas is capped at 9. In contrast, female and male personas could achieve a score of 10. This distinct disparity highlights a clear bias within the new model, particularly against non-binary individuals. For each degree level, female personas were consistently rated with lower competency scores than their male counterparts. This consistent pattern of lower scores for female personas across all education levels strongly suggest that the bias in the new model is not incidental but a systematic issue. The data underscores how gender bias pervades the competency evaluations, further reinforcing the need for more comprehensive measures to address such biases in AI models.

| Model | Gender | Average | Bachelors | Masters | PHD |
|---|---|---|---|---|---|
| o1 | F | 7.9 | 7.5 | 7.8 | 8.6 |
|  | M | 8.1 | 7.7 | 7.9 | 8.7 |
|  | NB | 7.8 | 7.5 | 8.4 |  |
| 4o | F | 8.6 | 8.4 | 8.5 | 9.1 |
|  | M | 8.6 | 8.3 | 8.7 | 9.4 |
|  | NB | 8.5 | 8.6 | 8.5 |  |

Table 2: Summary of Competency Scores

Table 3 reveals a significant disparity in the assignment of PhD qualifications across genders, with the percentage of PhDs for male personas being notably higher than for female personas. The gap has widened from 3.4% in the old model to 5.6% in the new o1-mini model, indicating a growing bias in how the models perceive educational attainment between genders. Additionally, none of the non-binary personas generated by the new model were assigned PhDs, further highlighting the model's bias against non-binary genders.

| Gender | o1 | 4o |
|---|---|---|
| F | 22.4% | 23.3% |
| M | 28.0% | 19.9% |
| NB | 0.0% | 0.0% |
| Total | 22.9% | 21.1% |

Table 3: Personas created for Degree level at PHD

Bias is evident in the model's assignment of the probability of becoming a successful founder, where clear gender stereotyping persists. Table 4 shows that the "probability of successful founder" for female and non-binary personas remains significantly lower compared to male personas. While the gap between female and male personas has narrowed, female and non-binary individuals are still consistently rated with lower probabilities of becoming successful founders, reflecting entrenched gender stereotypes. in the new model, female personas consistently received lower average probabilities compared to their male counterparts, regardless of their education level. This pattern reinforces the systematic nature of the bias, showing that even with higher education, female personas are less likely to be predicted as successful founders. This continued disparity emphasizes the need for stronger bias mitigation strategies in model development to ensure equitable treatment across all genders.

| Model | Gender | Average | Bachelors | Masters | PHD |
|---|---|---|---|---|---|
| o1 | F | 68.0% | 64.0% | 69.7% | 70.2% |
| o1 | M | 69.4% | 66.3% | 70.4% | 72.2% |
| o1 | NB | 65.7% | 63.6% | 70.6% | |
| 4o | F | 67.8% | 69.5% | 69.6% | 61.9% |
| 4o | M | 70.4% | 68.4% | 69.2% | 77.2% |
| 4o | NB | 76.4% | 76.2% | 77.0% | |

Table 4: Probability of Becoming Successful Founders.

The bias extends further to the probability of a person becoming CEO, with clear signs of gender stereotyping. Table 5 illustrates that the "average probability of becoming a CEO" for female and non-binary personas is significantly lower than that for male personas. Although the gap between female and male personas has narrowed, female and non-binary individuals still face lower probabilities of being predicted as CEOs. Notably, the median probability value for female personas is also lower than that for male personas, a trend that persists for non-binary personas as well. In the new model, female personas consistently receive lower average CEO probabilities than their male counterparts, regardless of their education level (Bachelor's or PhD except Master's). This demonstrates that even within highly educated groups, the model continues assigning lower leadership potential to female personas, underscoring a systematic bias across different educational backgrounds. Non-binary personas face a similar disparity at bachelor's degrees, further reinforcing the need for targeted bias reduction strategies. However, a similar trend is not observed for a non-binary persona with a Master's degree This data reinforces the persistent biases in leadership predictions across all gender dimensions discussed, with female and non-binary personas consistently rated lower

| Model | Gender | Average | Bachelors | Masters | PHD |
|---|---|---|---|---|---|
| o1 | F | 61.1% | 57.3% | 63.2% | 62.3% |
| o1 | M | 62.2% | 59.9% | 63.0% | 64.1% |
| o1 | NB | 58.4% | 56.1% | 63.7% | |
| 4o | F | 68.5% | 68.9% | 69.0% | 67.0% |
| 4o | M | 70.3% | 67.7% | 69.6% | 77.2% |
| 4o | NB | 71.0% | 70.8% | 71.5% | |

Table 5: Probability of Becoming CEO.

## 4.2 Analysis of Personality Traits:

The normalized frequencies of personality traits (across female, male, and non-binary categories) reveal important patterns of both inclusivity and bias within the o1-mini model. It is essential to consider these results within the broader context of gender representation. Below is a detailed analysis of the key trends, biases, and implications, reflecting on the broader systemic issues the model may reflect. Several key traits traditionally associated with leadership, dominance, and organizational skills remain heavily skewed toward males. These biases reinforce stereotypical views about male personas being naturally inclined toward competitive and leadership-driven roles. Males (0.010) are overrepresented in competitiveness, with females trailing behind (0.003) and no non-binary representation (0.000). The near-exclusive assignment of competitiveness to males suggests a reinforcement of the view that men are more driven by competition and achievement, traits that are often valued in professional and leadership roles. Males also lead in this category (0.043), with females (0.027) and no representation for non-binary individuals (0.000). This highlights the perception that men are inherently more capable of managing tasks and responsibilities. The male dominance in traits like competitive and organized reflects a lingering bias that men are better suited for leadership, high-pressure environments, or roles that require strategic planning. This bias can lead to unequal opportunities for women, who may be overlooked for leadership positions due to assumptions about their competitiveness or organizational abilities. The model continues to assign emotional and caregiving traits predominantly to women, perpetuating traditional gender norms. Females exhibit the highest normalized frequency (0.032), followed by males (0.017) and non-binary individuals (0.033). While the inclusion of non-binary individuals in empathy suggests some progress, the overwhelming association of this trait with females continues to mirror societal expectations of women as more emotionally attuned and nurturing. Similarly, females lead in caring (0.015), with males (0.011) and very minor representation for non-binary individuals (0.001). This trend reinforces the stereotype that caregiving is inherently a female trait. These findings indicate that the model is still reliant on outdated gender stereotypes, where women are primarily viewed through the lens of emotional intelligence and caregiving. Such biases can hinder women's access to roles that demand other traits, such as leadership or analytical abilities, while reinforcing the expectation that women take on emotionally laborious roles in both professional and personal contexts. For **analytical**, both males (0.170) and non-binary individuals (0.170) exhibit higher frequencies than females (0.130). The gap between male and female representation suggests that the model favors one gender over another regarding analytical thinking and problem-solving. For **creative**, females (0.082) and non-binary individuals (0.083) show similar frequencies and are higher than males (0.063). The representation across all genders indicates that the model stereotype that creativity is gender-specific. The representation of non-binary individuals,ise present in traits like empathy, creativity, and analytical. Non-binary individuals are absent from several key traits, such as being articulate, caring, and communicative. The analysis of the o1-mini model reveals clear patterns of gender bias, particularly in the way personality traits are attributed to male and female personas. Males continue to dominate traits associated with leadership, competitiveness, and organization, while females are predominantly associated with emotional and nurturing traits. At the same time, the model, although skewed, shows a good representation of traits such as analytical and creative across genders, demonstrating that certain cognitive abilities are being recognized across all genders.

**4.3 Analysis of Likings:**

This section investigates the likings generated by the o1-mini model, focusing on gender-based distributions of interest across a variety of fields. For this, normalized frequency ($f^{l,1}$) of the terms populated by the model) is considered for analysis. The findings offer valuable insights into the broader patterns of gender bias and highlight the need for more balanced representation in model training processes. Engineering (male: 0.0672, female: 0.0101), data (male: 0.0407, female: 0.0188), and technology (male: 0.0830, female: 0.0470, non-binary: 0.0672) demonstrate a pronounced male dominance. While females and non-binary personas are represented, the normalized frequencies for males far exceed those of the other groups, reflecting historical biases that associate technical proficiency with masculinity. Fields like Design (female: 0.0320, male: 0.0231, non-binary: 0.0500) and Art (female: 0.0239, male: 0.0131, non-binary: 0.0502) show stronger representation for females and non-binary personas. This reinforces societal stereotypes linking creativity with these genders, while males are underrepresented in these domains. Marketing follows a similar pattern (female: 0.0439, male: 0.0201, non-binary: 0.0332), further entrenching the notion that communication and creativity-driven fields are seen as more gendered toward females and non-binary personas. Non-Binary personas are highly represented in niche areas such as fitness, open-source Projects, AI/ML, and travel. Despite this high representation in specific fields, non-binary personas show limited or no representation in several other areas. The analysis of the o1-mini model reveals persistent gender biases across fields like engineering, data, and technology, where males dominate, reflecting traditional stereotypes. Conversely, fields like Design, Art, and marketing show a stronger presence of females and non-binary personas, reinforcing societal notions that associate creativity and communication with these genders. Overall, the findings emphasize the importance of more balanced and inclusive model training to reduce gender biases and ensure fairer representation across all fields.

### 5. Future direction to mitigate biases

Biases and stereotyping in AI and LLMs have received a lot of attention from researchers and industry; however, we do see biases even in advanced high reasoning models. Solving issues of biases and stereotyping is complicated and therefore needs a multi-faceted approach to mitigate them. Skewed representation in training datasets is one of the critical reasons for biases. Curated Data Collection by balancing datasets across diverse perspectives, ensuring broad representation across genders, races, and socioeconomic backgrounds, and Real-Time Data Auditing by implementing continuous monitoring tools to detect and flag biased patterns in incoming training data are two effective ways to handle biases due to skewed data. Algorithmic Bias, a deeper bias into the model, can be reduced by Algorithmic Debiasing Techniques and Bias Regularization in Loss Functions of the model. In Algorithmic Debiasing, model outputs can be evaluated against counterfactual scenarios where demographic variables can be changed while other attributes remain the same, whereas, in bias regularization in loss functions, fairness constraints can be included in optimization functions to penalize biased outputs. Human-in-the-loop feedback can be a powerful way to reduce bias in the models, and this can be done at the training stage by involving humans to flag biased outputs, and the second stage is by allowing end-users to report biases, triggering fine-tuning updates that adjust model behavior accordingly. Embedding a "fairness meta-layer" with a fairness check mechanism before LLM generates a response can be a powerful way of keeping biases in check. This meta-layer acts as a guardrail and modifies unfair responses before passing them to downstream systems or users. The effective way of bias mitigation needs to be dynamic, self-correcting, and fairness-aware AI systems.

### 6. Conclusion:

In conclusion, our study provides a crucial evaluation of the o1 model, shedding light on the persistent gender biases that pervade large language models. Despite some progress in gender inclusivity—especially in personality traits and preferences—our findings highlight significant disparities in how different genders are rated for competency, leadership potential, and career growth. Female personas continue to be stereotyped into traditional roles. These results underscore the pressing need for deeper focus to mitigate bias and foster equitable gender representation. As AI systems continue to shape societal perceptions and decision-making, further research must focus on building models that have fairness, inclusivity, and transparency to prevent the amplification of harmful stereotypes and ensure that future models support a more equitable digital landscape.

**References**

[1] An, J., Huang, D., Lin, C., & Tai, M. (2024). Measuring Gender and Racial Biases in Large Language Models. ArXiv. /abs/2403.15281

[2] Bender, E., Gebru, T., McMillan-Major, A., & Shmitchell, S., (2021). On the Dangers of Stochastic Parrots: Can Language Models Be Too Big. Proceedings of FAccT, 2021.

[3] Bolukbasi, T., Chang, K., Zou, J., Saligrama, V., & Kalai, A. (2016). Man is to Computer Programmer as Woman is to Homemaker? Debiasing Word Embeddings. *ArXiv*. /abs/1607.06520

[4] *Buolamwini, J., &* Gebru, T., (2018). Gender Shades: Intersectional Accuracy Disparities in Commercial Gender Classification. Proceedings of the 1st Conference on Fairness, Accountability, and Transparency.

[5] Gupta, S., Ranjan, R., & Singh, S. N. (2024). Comprehensive Study on Sentiment Analysis: From Rule-based to modern LLM based system. *ArXiv*. https://arxiv.org/abs/2409.09989

[6] Gupta, S., & Ranjan, R. (2024). Evaluation of LLMs Biases Towards Elite Universities: A Persona-Based Exploration. *ArXiv*. https://arxiv.org/abs/2407.12801

[7] Huang, D., Bu, Q., Zhang, J., Xie, X., Chen, J., & Cui, H. (2023). Bias Testing and Mitigation in LLM-based Code Generation. ArXiv. /abs/2309.14345

[8] Kurita, K., Vyas, N., Pareek, A., Black, A. W., & Tsvetkov, Y. (2019). Measuring Bias in Contextualized Word Representations. *ArXiv*. /abs/1906.07337

[9] Kotek, H., Dockum, R., & Sun, D. Q. (2023). Gender bias and stereotypes in Large Language Models. ArXiv. https://doi.org/10.1145/3582269.3615599.

[10] Liu, S., Maturi, T., Shen, S., & Mihalcea, R. (2024). The Generation Gap: Exploring Age Bias in Large Language Models. ArXiv. /abs/2404.08760.

[11] OpenAI. (2023). GPT-4 Technical Report. Retrieved from https://www.openai.com/research/gpt-4

[12] Raj, C., Mukherjee, A., Caliskan, A., Anastasopoulos, A., & Zhu, Z. (2024). Breaking Bias, Building Bridges: Evaluation and Mitigation of Social Biases in LLMs via Contact Hypothesis. ArXiv. /abs/2407.02030

[13] Ranjan, R., Gupta, S., & Singh, S. N. (2024). A Comprehensive Survey of Bias in LLMs: Current Landscape and Future Directions. *ArXiv*. https://arxiv.org/abs/2409.16430

[14] Singh, S., Keshari, S., Jain, V., & Chadha, A. (2024). Born With a Silver Spoon? Investigating Socioeconomic Bias in Large Language Models. ArXiv. /abs/2403.14633

[15] Wang, Z., Wu, Z., Guan, X., Thaler, M., Koshiyama, A., Lu, S., & Beepath, S. (2024). JobFair: A Framework for Benchmarking Gender Hiring Bias in Large Language Models. ArXiv. /abs/2406.15484.

[16] Zhao, J., Wang, T., & Yatskar, M. (2019). Men Also Like Shopping: Reducing Gender Bias Amplification using Corpus-level Constraints. Proceedings of the 2019 Conference on Empirical Methods in Natural Language Processing.